\documentstyle[aps,tighten,epsf]{revtex}
\begin{document}

\title{Quantization of Hall Conductance in Double Exchange Systems:\\
Topology and Lattice Gauge Field}

\author{Atsuo Satou and Masanori Yamanaka}

\address{Department of Applied Physics, Science University of Tokyo, 
Kagurasaka, Shinjuku-ku, Tokyo 162-8601, Japan}


\maketitle

\begin{abstract}
We study quantization conditions of the Hall conductivity 
for a two dimensional system described by a 
double exchange Hamiltonian
with and without an external magnetic field.
This is obtained by an extension of the topological arguments
familiar from the theory of the integer quantum Hall effect. 
The quantization conditions are related to spontaneous breaking 
of spin $O(3)$, time-reversal, and spin chiral symmetries.
Extension to systems with higher dimensions is briefly discussed.
\end{abstract}


The quantum Hall effect (QHE) is a 
remarkable phenomena in condensed matter physics \cite{REFPrange}. 
After its discovery, there have been many theoretical works 
which explained the quantization of the Hall 
conductivity $\sigma_{xy}= n \frac {e^2} {h}$ 
for two dimensional electron gas in a uniform 
perpendicular magnetic field. An elegant approach, 
of which we will make use below, is 
based on topological and symmetry arguments.
For Bloch electrons in a magnetic field,
several authors,~\cite{REFTKNN,REFkohmotoann}
derived an explicit formula for the Hall conductance 
which is independent on the detailed structure
of the periodic potential~\cite{REFdana}.
The integer $n$ is shown to be a topological invariant,
the first Chern class of a $U(1)$ principal fiber bundle
on a torus~\cite{REFkohmotoann}. 
(See also~\cite{REFnovi,REFAvron}.)
In a pioneering work by Haldane~\cite{REFHaldane},
the possible occurrence of
QHE without an external magnetic field
in systems with broken time-reversal symmetry
and its relation to chiral anomaly 
in 2d field theories have been discussed.
(See also~\cite{REFwiegmann}.)

The QHE is mainly concerned with 
charge degrees of freedom of electrons. 
An experimentally and theoretically 
relevant question is, therefore, whether there is 
a similar effect pertaining to the spin degrees of freedom.  
Indeed, in transition metal oxides,
coupling of spin degrees of freedom 
between itinerant and localized electrons 
induces many interesting transport properties~\cite{REFtokura}.

In this letter we show that, in principle, a special                   
variant of the QHE may occur also in transition metal oxides
with and without an external magnetic field.
As we have reported recently~\cite{REFYKM},  
the basic ingredients are the Double Exchange (DE) model~\cite{REFzener}
and the theory developed in Refs.~\cite{REFTKNN,REFkohmotoann}.
The DE model
exhibits rich transport properties~\cite{REFkubo,REFtokura}.
The phase factor in the hopping integral~\cite{REFmuller}
induced by the $t_{2g}$ spins 
leads to an exotic ground state,
(referred to as ``flux'' state),
where the Hall conductance is expected 
to be quantized~\cite{REFYKM,REFomn}.
The mechanism of stabilization of the ``flux'' state
is similar to that of the
flux phase discussed in the Hubbard 
model~\cite{REFam,REFLieb} 
and in a generalized Peierls instability~\cite{REFHasegawa}.
Numerous other facets due to the Berry phase
or to the ``flux'' were studied, 
such as an anomalous Hall effect ~\cite{REFYBK,REFomn},
Jhan-Teller effect~\cite{REFkoizumi}, 
and stripe formation~\cite{REFhotta}.

The conditions we study for the quantization 
of the Hall conductivity are summarized as follows:
(i) the Fermi level should lie in a gap of extended states~\cite{REFTKNN}.
(ii) the ground state is not degenerate~\cite{REFTKNN}.
(iii) time-reversal symmetry is broken\cite{REFHaldane}.
Condition (iii) is evidently realized in the presence 
of a uniform external magnetic field in the standard QHE.
Alternatively, in the DE model one may perceive two
possibilities of breaking time-reversal symmetry:
(a) Due to the minimum principle of external flux force line,
it is natural to conjecture that the ground state of the DE model 
has a staggered flux structure~\cite{REFYKM}
where the net spontaneous flux vanishes.
However, if a state with non-vanishing spontaneous flux exists,
it breaks time-reversal symmetry.
The pertinent state is characterized
by the Wilson loop~\cite{REFwilson}.
The mechanism of stabilization of such a state was discussed 
in Refs~\cite{REFYBK,REFomn}.
(b) Even if  case (a) is hardly realizable,
a system with a proper lattice structure 
spontaneously breaks the time-reversal symmetry.
This kind of symmetry breaking corresponds to 
that of spin chirality~\cite{REFYBK,REFomn}. 
One such example is the Kagome lattice~\cite{REFomn}.
In the rest of this letter,
we first study the Hall conductivity in its general form
within the DE model along the lines of TKNN, 
and then consider some specific examples.
Finally, extension to systems with higher dimensions 
and the relation with theorems due to Lieb~\cite{REFLieb}
and Elitzur~\cite{REFelitzur} are briefly discussed.

The Hamiltonian of
the DE model in its minimum version~\cite{REFcom4}
at finite doping reads,
\begin{eqnarray}
H= \sum_{i,j;\mu}
   \Big( -t c^{\dagger}_{i,\mu}c^{\phantom{\dagger}}_{j,\mu}
  + h.c.
  + J \vec{S}_i \cdot \vec{S}_j \Big)
  - J_H \sum_i \vec{\sigma}_i \cdot \vec{S}_i,
\label{eq:hamiltonian}
\end{eqnarray}
where $c_{i,\mu}$ is a fermion 
(in fact, an $e_g$ electron) annihilation operator 
at site $i$ with spin projection $\mu$, 
$\vec{\sigma}$ is the spin operator of the $e_g$ electrons, 
$\vec{S}_i$ are localized ($t_{2g}$) spins 
which are treated as classical vectors directed along
$(\theta_i, \phi_i)$ in spherical coordinates.
Moreover, $J_H$ is the Hund rule coupling constant and 
$J$ is the direct exchange coupling strength between $t_{2g}$ spins.
There is a local $SU$(2) rotation invariance at every site,
and, in the limit $J_H \to \infty$, 
the original Hamiltonian transforms into~\cite{REFlocalSU2},
\begin{eqnarray} 
\tilde{H}=-t \sum_{\langle i, j \rangle} 
\Big[
\Big(
\cos{\frac{\theta_i}{2}}
\cos{\frac{\theta_j}{2}}
+ e^{-i ( \phi_i - \phi_j)}
\sin{\frac{\theta_i}{2}}
\sin{\frac{\theta_j}{2}}
\Big) 
c^{\dagger}_i c^{\phantom{\dagger}}_j + h.c.
\Big] 
+J 
\sum_{\langle i, j \rangle} 
\vec{S}_i \cdot \vec{S}_j.
\label{eq:effectiveham}
\end{eqnarray} 
The current operator on link $ij$  
is transformed as
\begin{eqnarray}
v_{ij}=\frac{i}{\hbar}
\big(c^{\dagger}_i c^{\phantom{\dagger}}_j
-c^{\dagger}_j c^{\phantom{\dagger}}_i \big) 
\to
\tilde{v}_{ij}=\frac{i}{\hbar}
\Big[\cos{\frac{\theta_i}{2}}\cos{\frac{\theta_j}{2}}
\big(c^{\dagger}_i c^{\phantom{\dagger}}_j
-c^{\dagger}_j c^{\phantom{\dagger}}_i \big)
+
\sin{\frac{\theta_i}{2}}
\sin{\frac{\theta_j}{2}}
\big(c^{\dagger}_i c^{\phantom{\dagger}}_j
 e^{-i ( \phi_i - \phi_j)}
- 
c^{\dagger}_j c^{\phantom{\dagger}}_i
 e^{i ( \phi_i - \phi_j)}\big)\Big].
\label{eq:currentop}
\end{eqnarray}
The transformation is unitary and 
does not affect the energy spectrum.
However, the link variable which is employed to calculate 
the Hall conductance is not invariant.
Below, we calculate the Hall conductacne, $\sigma_{xy}$,
associated with the Hamiltonian (\ref{eq:effectiveham}).
The phase factor in the second term in (\ref{eq:currentop}) is 
induced from the $t_{2g}$ spin configuration, which we
denote by $\{\theta_i, \phi_i\}$ below. 
The structure of the phase factor is related 
to the non-trivial magnetic ordering pattern 
$\{\theta_i, \phi_i\}$.
In a generic configuration, 
the $e_g$ electron acquires flux 
which cannot be eliminated by a local gauge transformation,
that is, it accumulates a non-zero phase (mod $2\pi$) 
on moving around a closed path.
One of the realizations of this scenario  
is the ``flux'' state~\cite{REFYKM}.

We now explain the calculation of $\sigma_{xy}$. 
In the absence of an external magnetic field,
the contribution from the first term in (\ref{eq:currentop}) vanishes,
so let us concentrate on the second one.
(When the field is switched on, the conductivity is
the sum of the two terms.
The contribution from the first term is 
obtained within the TKNN formalism.)
In the course of its evaluation, 
one has to employ a Fourier transformation.
However, this turn out to be useless
since in generic situations, the spin configuration $\{\theta_i, \phi_i\}$
is not periodic.
This is due to the property of the DE model 
where the energy is a functional of the {\em oriented relative angle} 
of the $t_{2g}$ spin~\cite{REFYKM,REFKYOM}.
Thus, for a generic spin configuration,
there is no simple algorithm for this calculation problem.
Below, we assume a periodicity of the $\{\theta_i, \phi_i\}$ configuration.
This can be justified 
at least for the staggered flux state~\cite{REFYKM}
by noting 
that the flux state is degenerate.
We define the respective unit cells for
$\theta$'s and $\phi$'s, 
$U_{\theta}$ and $U_{\phi}$, 
which are not necessarily identical.
These unit cells can be enlarged to the least common multiple 
(denoted hereafter by $U$) of
$U_{\theta}$ and $U_{\phi}$, if their ratio is rational. 
(An interesting situation occurs 
when the ratio of $U_{\theta}$ and $U_{\phi}$ is irrational.)

The expression for the Hall conductivity now reads \cite{REFkohmotoann},
\begin{eqnarray}
\sigma_{xy} = \sum_j\bigg[\frac{e^2}{h}\frac{1}{2\pi i}\oint_{MBZ}d\vec{k}\cdot
\langle \vec{k} \vert \nabla_k \vert \vec{k} \rangle \bigg]_j
=\frac{e^2}{h} t_r,
\label{eq:integral}
\end{eqnarray}
where $\vec{k}$ is the crystal momentum and
$\vert \vec{k} \rangle$ is a wave function
in momentum space defined on the extended magnetic Brillouin zone 
(MBZ) associated with $U$.
Note that here, the MBZ refers
to the Brillouin zone of the ``spin configuration.''
Here $j$ is the index of the sub-band.
Finally, the number $t_r$ on the right hand side 
of equation (\ref{eq:integral})
is the solution of the Diophantine equation~\cite{REFTKNN}.

The simplest example for which the above formalism 
can be worked out in detail is
the DE model on a square lattice at half-filling.
The staggered $\pi$-flux state is 
spontaneously stabilized~\cite{REFYKM},
and the ground state has an infinite continuous degeneracy.
The dispersion relation near the Fermi level is approximated
by that for a gapless Dirac fermion.
In the absence of an external magnetic field,
the time-reversal symmetry is not broken
because the spontaneous flux around a plaquette is $\pm\pi$,
and the Hall conductance vanishes.
Below we assume that 
the symmetry of the $t_{2g}$ spin is broken, 
thus removing the continuous degeneracy. We also introduce 
a next-nearest neighbor (NNN) hopping term for staggared plaquette
in order to open an energy gap~\cite{REFHK1} 
and to break time-reversal symmetry~\cite{REFHaldane}.
We further assume that the staggered structure of the $\pi$-flux
in (\ref{eq:effectiveham}) is robust 
against perturbation by NNN hopping.
The effective Hamiltonian for the $e_g$ electron is described by 
\begin{eqnarray} 
H&=&-t \sum_{m,n} \Big[
e^{i\theta_1}
c^{\dagger}_{m+\frac{1}{2},n+\frac{1}{2}} c^{\phantom{\dagger}}_{m,n} 
+e^{i\theta_2}
c^{\dagger}_{m+\frac{1}{2},n+\frac{1}{2}} c^{\phantom{\dagger}}_{m+1,n}
+e^{i\theta_3}
c^{\dagger}_{m,n+1} c^{\phantom{\dagger}}_{m+\frac{1}{2},n+\frac{1}{2}}
\nonumber\\
&&\hspace{12mm}
+e^{i\theta_2}
c^{\dagger}_{m+1,n+1} c^{\phantom{\dagger}}_{m+\frac{1}{2},n+\frac{1}{2}}
-t' \Big(
e^{i\theta_4}
c^{\dagger}_{m,n+1} c^{\phantom{\dagger}}_{m,n}
+e^{i \pi \varphi}
c^{\dagger}_{m+\frac{1}{2},n+\frac{1}{2}} 
c^{\phantom{\dagger}}_{m-\frac{1}{2},n+\frac{1}{2}}
\Big) \Big] + h.c.,
\label{eq:staggaredham}
\end{eqnarray} 
where $\theta_1=\pi[\phi(m+\frac{1}{4})+i2\varphi]$,
$\theta_2=\pi\phi(m+\frac{3}{4})$,
$\theta_3=\pi\phi(m+\frac{1}{4})$,
and $\theta_4=2\pi\phi m+i \pi \varphi$.
Here the effect of the spontaneous staggered $\pi$-flux 
is replaced by $\varphi=1/2$.
Recently, it has been possible to fabricate
networks which are described by 
the Hamiltonian (\ref{eq:staggaredham}) with $t'=0$
and to investigate the corresponding
energy spectrum~\cite{REFandoiye}. In our calculations,
we introduce also a uniform external magnetic field 
$\phi$ (in the Landau gauge) thus taking into account 
the experimental setup.

We now derive the Chern number using the 
method detailed in Ref.~\cite{REFHatsugai}
which provides a simple counting technique for the TKNN-vortex.
In the absence of the external field, i.e. $\phi=0$,
the Hall conductance is quantized as a function of 
the perturbation, $t_r=sgn(t')$ and $t_r=0$ for $t'=0$.
The value of $t_r$ for finite 
$\phi$ is shown in Fig.~\ref{FIGUREsigmaxy}
where the staggered flux state of the DE model
can be interpreted only near $\phi \sim 0$.

In the above formalism, we artificially break 
the time-reversal symmetry
and open the energy gap. 
However, the symmetry would be spontaneously broken
away from half-filling
because the super-cell structure is expected to stabilize
if one employs the analogy between the present 
model and the generalized 
Peierls instability~\cite{REFHasegawa}. 

The geometrical structure of the lattice 
may induce a non-trivial spin configuration~\cite{REFomn}.
Motivated by the work
we study the DE model on a two dimensional pyrochlore 
like lattice~\cite{REFkagome},
defined by assigning tetrahedrons on the Kagome lattice
shared with one of the four triangles 
(see Fig.\ref{FIGUREpylochrore}(a)).
In the limit $J \to \infty$, 
the ground state has continuous infinite degeneracy.
(Its nature is distinct from 
that of the Kagome lattice~\cite{REFkagome} 
where the relative angle 
between adjacent spins is $2/3\pi$.
In the pyrochlore like lattice,
the corresponding angles
need not be identical.
Typical configurations are displayed 
in Fig.\ref{FIGUREpylochrore}(b)-(e).
The spin structure over entire lattice is obtained 
by combining these tetrahedrons and successively
adjusting the relative angles of the shared edges.)
Because of the degeneracy,
one cannot evaluate (\ref{eq:integral}).
Below we assume that one of the spin configurations is
selected due to symmetry breaking of the spin
and that the periodicity coincides with
the Wigner-Seitz cell of the lattice.
Employing the method of Ref.~\cite{REFYKM,REFKYOM}
we optimize the four spin configuration in the Wigner-Seitz cell
in order to minimize the kinetic and exchange energies.
We studied the model at fillings
$x=$ 1/5, 2/5, and 3/5 and confirmed the above picture.
Two simple and symmetric realizations are shown 
in Figs.\ref{FIGUREpylochrore}(f) and (g).
In these examples, the flux through a triangle is exactly $\pi$
and the time-reversal symmetry is not broken.
Note that in the degenerate ground state, 
there exists a spin configuration
with broken time-reversal symmetry. 
An example is shown in Fig.\ref{FIGUREpylochrore}(e).
However, here we study the most extreme cases (f) and (g).

The effective Hamiltonian associated with these states is
\begin{eqnarray} 
H=-t \sum_{m,n} \Big[&&
e^{i\frac{2}{3}\pi|\varphi|}
c^{\dagger}_{m+1,n} c^{\phantom{\dagger}}_{m,n} 
+e^{-i\frac{2}{3}\pi\varphi}
c^{\dagger}_{m,n} c^{\phantom{\dagger}}_{m-1,n}
+e^{i2\pi m\phi-i\frac{2}{3}\pi|\varphi|}
c^{\dagger}_{m,n+1} c^{\phantom{\dagger}}_{m,n}
+e^{i2\pi m\phi+i\frac{2}{3}\pi\varphi}
c^{\dagger}_{m,n} c^{\phantom{\dagger}}_{m,n-1}
\nonumber\\
& &
+c^{(1)\dagger}_{m+1,n+1} c^{\phantom{\dagger}}_{m,n}
+c^{\dagger}_{m,n} c^{(2)}_{m-1,n-1}
+e^{i(2m+\frac{1}{3})\pi\phi}
c^{(1)\dagger}_{m+1,n+1} c^{\phantom{\dagger}}_{m,n+1}
+e^{-i(2m-\frac{1}{3})\pi\phi}
c^{\dagger}_{m,n-1} c^{(2)}_{m-1,n-1}
\nonumber\\
& &
+e^{i\frac{1}{3}\phi\pi}
c^{(1)\dagger}_{m+1,n+1} c^{\phantom{\dagger}}_{m+1,n}
+e^{-i\frac{1}{3}\phi\pi}
c^{\dagger}_{m-1,n} c^{(2)}_{m-1,n-1}
\nonumber\\
& &
+e^{i\pi\phi(2m+1)+i\frac{2}{3}\pi|\varphi|}
c^{\dagger}_{m,n+1} c^{\phantom{\dagger}}_{m+1,n}
+e^{i\pi\phi(2m-1)-i\frac{2}{3}\pi\varphi}
c^{\dagger}_{m-1,n} c^{\phantom{\dagger}}_{m,n-1}
\Big] + h.c.
\label{eq:pylochore}
\end{eqnarray} 
Here the effect of the spontaneous staggered $\pi$-flux 
is replaced by $\varphi=\pm 1/2$ 
for Figs.\ref{FIGUREpylochrore}(f) and (g), respectively.
We also introduce a uniform external magnetic field 
$\phi$ in the Landau gauge and study the energy gap structure
of the model.
For a generic $\varphi$, the state shown 
in Fig.\ref{FIGUREpylochrore}(f) breaks time-reversal, 
while the state shown in (g) does not.
These two states with different behavior under time reversal
are (Kramer) degenerate.

The pertinent Chern number 
is calculated by using the method developed 
in Refs.~\cite{REFHatsugai,REFyhk}
where, in the present case,
the structure of the complex energy surface 
is classified as a Riemann surface 
associated with a ten degree algebraic equation 
which we have solved numerically.
The $t_r$'s are shown in Fig.\ref{FIGUREsigmaxy}(b).
The region $\phi \sim 0$ corresponds to 
the states depicted in Fig.\ref{FIGUREpylochrore}(f) and (g).
Perturbation by additional hopping matrix elements
to the honeycomb cell (which is identical to the one 
discussed in Ref.~\cite{REFHaldane}) opens an energy gap
and $\sigma_{xy}$ is quantized to the value 
associated with the nearest energy gap.

Realizing the quantization conditions (i-iii) 
is most difficult for the states which we have studied above.
Introduction of an anisotropy in the hopping amplitude
or exchange energy between in and out of plane bonds
induces a tilt from the $\pi$-flux,
and the time-reversal symmetry is broken.  
As another approach, we can construct 
the entire spin configuration
from that of the tetrahedron configuration
with broken time-reversal symmetry
using the unit shown in Fig.\ref{FIGUREpylochrore}(e).
The occurrence of the QHE in 3$d$ pyrochlore 
was speculated~\cite{REFomn}.
We confirmed that the energy gap G in Fig.~\ref{FIGUREsigmaxy}(b)
survives in 3$d$ pyrochlore 
for an anisotropic inter-layer hopping integral.
This leads the QHE in 3$d$~\cite{REFhalperin,REFkhw,REF3dcond}.

{\em Discussion:}
We have investigated the quantization condition of 
$\sigma_{xy}$ in the ground state of the DE model.
In general, it is hard to simultaneously maintain
the conditions (i-iii).
Especially, the ground state of the DE model 
have a continuous infinite degeneracy~\cite{REFKYOM}
which violates the condition (ii).
(The Kramer degeneracy also violates it.)
Some different degenerate spin configurations 
lead the same band structure of the $e_g$ electrons,
and the other spin configurations might lead to different 
band structures.
(We denote the Bloch wave associated with the band structures
by $\vert k_i \rangle$ where $i$ is the index of degeneracy.)
The full wave function is a superposition of Bloch functions
and the integral in equation (\ref{eq:integral}) is ill defined
except when
all $\vert k_i \rangle$ have the same Chern number.
{\em Therefore, we should stress that 
in most cases, the Hall conductance either vanishes
or it is ill defined,
even when the Fermi energy lies in an energy gap.}
The problem associated with this type of degeneracy 
is not explicitly addressed in TKNN.

On the other hand, the possibility of the quantization 
cannot be ruled out if symmetry breaking occurs.
The {\em global} spin O(3) symmetry breaking is allowed 
at zero temperature in 2$d$ or at finite temperature in 3$d$. 
The Chern number is different for each 
spin configuration and the quantized Hall conductivity 
depends on spin configuration.
Therefore, the value of the quantization 
may serve as a probe determining the spin configuration.
(In three or higher dimensions,
the QHE occurs~\cite{REFhalperin,REFkhw,REF3dcond}
in transition metal oxides 
if the system satisfies the conditions (i-iii).
For 3$d$, each band has three topological invariants
(the first Chern numbers) on a 2-tori obtained 
by slicing the three-torus in three different manners.
For general $d$, every quantized invariant on a $d$-dimensional torus 
$T^d$ is a function of the $d(d-1)/2$ sets of TKNN integers
obtained by slicing $T^d$ by the $d(d-1)/2$ distinct
$T^2$~\cite{REFAvron}.)

The DE model with classical spin has {\it local} symmetry
because the ground state energy is a functional of the 
{\em oriented relative angles} between spins.
The symmetry breaking is difficult.
Its relation to the Elitzur's theorem~\cite{REFelitzur}
is quite interesting.
In this case, the Hall conductance again vanishes or undefined. 
At least, at half-filling, the $\pi$--flux state is expected 
to be stabilized in the DE model with quantum spin
(even in three dimensions and/or with interactions),
because the system maintains
reflection positivity in spin space~\cite{REFLieb}.

In closing, we point out that flux states in the DE model 
manifest an interesting physics associated with 
the QHE and transport properties of
transition metal oxides, as well as of
networks with modulated nano-structures~\cite{REFandoiye}.

\noindent
{\bf Acknowledgment}: We would like to thank Y.~Avishai and M.~Oshikawa 
for stimulating discussions and comments on the manuscript.
The authors thank the Supercomputer Center, ISSP, 
U.Tokyo for the use of the facilities.
M.Y. is supported by the Moritani scholarship foundation.

\vspace{90mm}

\begin{figure}
\begin{center}
\caption{
Band structures (a) and (b) as a function of 
$\phi$ for the Hamiltonians 
(\ref{eq:staggaredham}) and (\ref{eq:pylochore}), respectively.
Numbers in the energy gaps are $t_r$ in (\ref{eq:integral}).
}
\label{FIGUREsigmaxy}
\end{center}
\end{figure}

\begin{figure}
\begin{center}
\caption{
(a) Two dimensional pyrochlore like lattice.
(b-e) Optimized spin configurations 
for single tetrahedron cluster in $J\to\infty$.
((b-d) The relative angles are the same 
and the flux penetrating a triangle is $\pi$.
(e) The relative angles are not the same
and the flux penetrating a triangle is not $\pi$.)
(f) flux configurations constructed from (b) and (c).
(g) flux configurations constructed (d), and (b) or (c).
}
\label{FIGUREpylochrore}
\end{center}
\end{figure}


\begin{references}
\bibitem{REFcom4}
Inclusion of orbital or other degrees of freedom 
is straightforward.
\end{references}
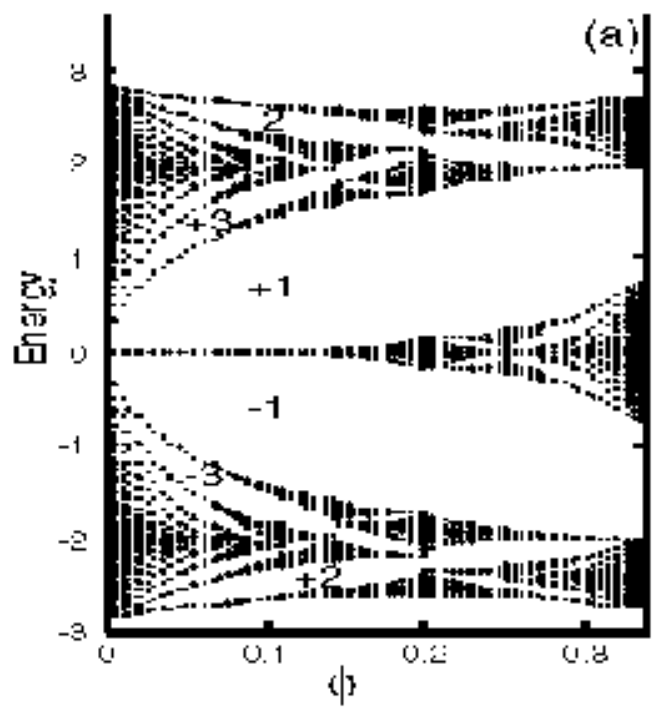
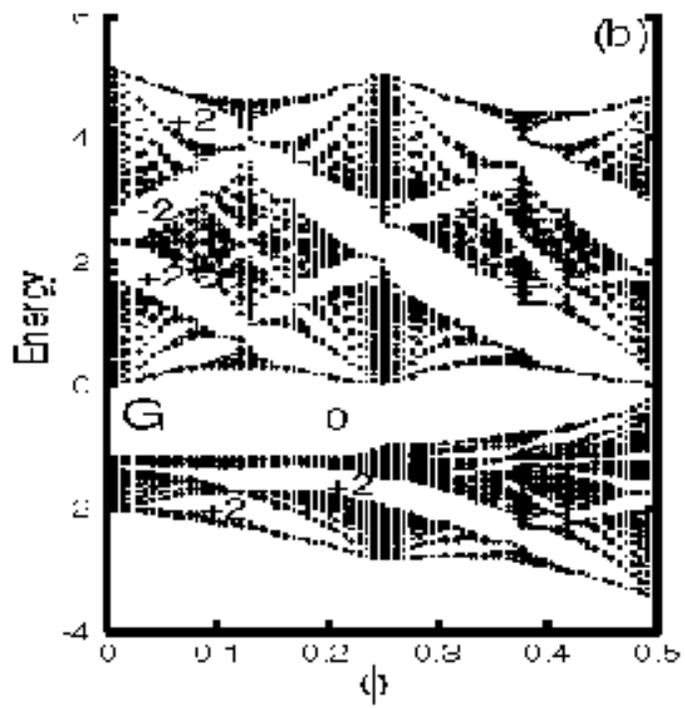
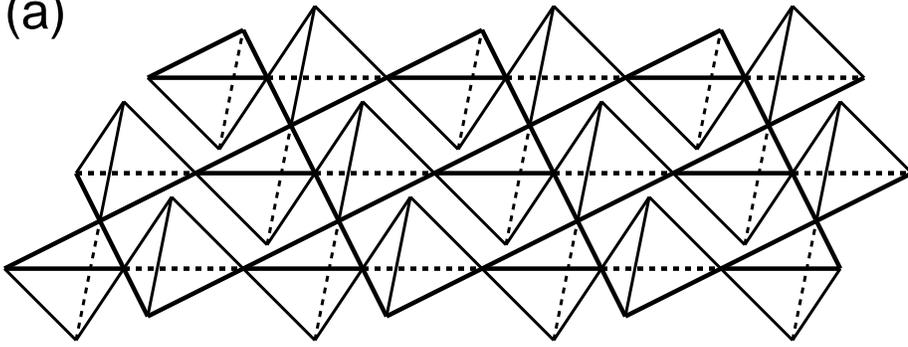
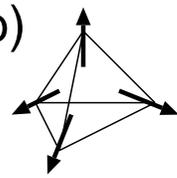
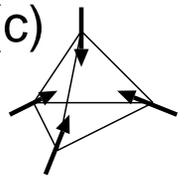
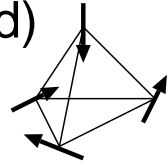
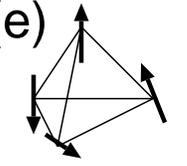
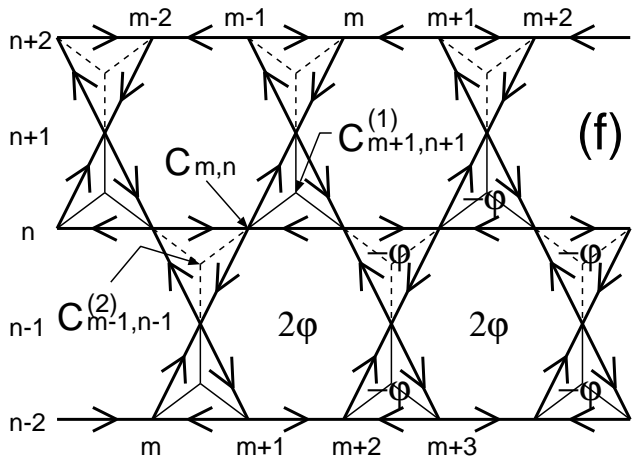
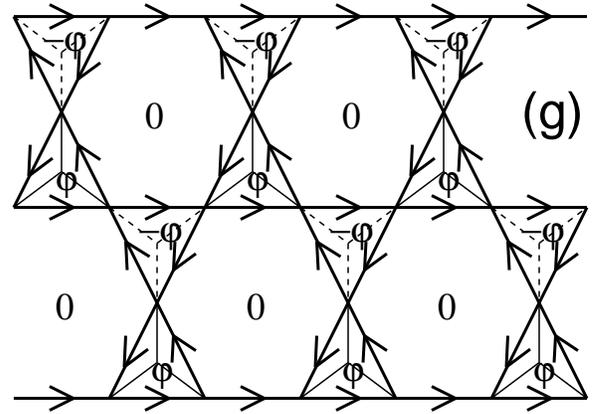
\end{document}